\documentclass[11pt, twoside]{paper}
\usepackage[german,english]{babel}
\usepackage{a4wide}
\usepackage{subfigure}
\usepackage{graphicx}
\usepackage{epsfig}
\usepackage{paralist}
\usepackage[usenames]{color}
\usepackage{booktabs}
\usepackage {eurosym}
\usepackage {lscape}
\usepackage{scalefnt}
\usepackage{enumitem}

\begin{document}
\title{Latest results of the direct dark matter search with the EDELWEISS-II experiment}

\author{{\slshape Gilles Gerbier $^1$ for EDELWEISS collaboration}\\[1ex]
$^1$IRFU/SPP, CEA Saclay, , 91191 Gif s Yvette, France }




\maketitle

\begin{abstract}
The EDELWEISS-II experiment uses cryogenic heat-and-ionization Germanium detectors in order to detect the rare interactions from WIMP dark matter particles of local halo. New-generation detectors with an interleaved electrode geometry were developped and validated, enabling an outstanding gamma-ray and surface interaction rejection. We present here preliminary results of a one-year WIMP search carried out with 10 of such detectors in the Laboratoire Souterrain de Modane. A sensitivity to the spin-independent WIMP-nucleon cross-section of 5$\times$10$^{-8}$ pb was achieved using a 322 kg.days effective exposure. We also present the current status of the experiment and prospects to improve the present sensitivity by an order of magnitude in the near future.
\end{abstract}

\section{The Edelweiss II set up and the detectors}

Weakly Interacting Massive Particles (WIMPs) are  a class of particles  now widely considered as one of the most likely  explanation for the various observations of dark matter from the largest scales of the Universe to galactic scales. The collisions of WIMPs on ordinary matter are expected to generate mostly  elastic scatters off nuclei, characterised  by low-energy deposits ($<$100 keV) with an exponential-like spectrum, and a very low interaction rate, currently constrained at the level of 1 event/kg/year. Detecting these events requires an ultralow radioactivity environment as well as detectors with a low energy threshold and an active rejection of the residual backgrounds~\cite{Bertone_book:2010}.

EDELWEISS-II is a low-background experiment using cryogenic Germanium detectors aiming at the direct observation of the local WIMPs which may constitute the dark matter of our Milky Way. 
The main EDELWEISS-II setup is located at the Laboratoire Souterrain de Modane (LSM), where the 4800 m water-equivalent rock overburden reduces the cosmic muon ßux down to 5~$\mu$/m$^2$/day. Germanium detectors are hosted within a reversed geometry dilution cryostat which may host up to 40 kg of detectors down to 20~mK. A 20 cm thick lead shield surrounds the detectors, with its inner parts made of roman lead, in order to attenuate the external $\gamma/\beta$ radioactivity. A 50 cm thick polyethylen shielding protects detectors against the external flux of fast neutrons. A muon veto made of plastic scintillators with a coverage of more than $>$98\% tags neutrons produced by the residual flux of muons which have been interacting mostly in the lead shield. Additional background monitorings are achieved by a Radon level detector near the cryostat, measurements of the thermal neutron flux inside the shielding with a $^{3}$He-gas detector, and studies of muon-induced neutrons with a Gd-loaded liquid scintillator outside the shielding. 

EDELWEISS detectors are ultrapure germanium crystals equipped with a dual heat and ionization measurement in order to discriminate induced electronic recoils from potential WIMP induced nuclear recoils, a technology with proven rejection efÞciency since the beginning of the 2000s ~\cite{Benoit:2001}. The heat sensors are NTD thermometers glued on the surface of each detector, while the ionization is measured using electrodes polarized at a few volts. The ionization yield for nuclear recoils is well-measured and is a factor $\sim$3 lower with respect to electron recoils in the energy region of interest, enabling a complete event-by-event discrimination for the bulk of $\gamma$-ray radioactivity. However, when an interaction takes place near the surface of the detector, the charge collection is incomplete and difÞcult to control. In particular, the local  radioactivity from residual $^{210}$Pb, a daughter of Radon which is present on all surfaces, generates such surface events. Deficit of charge collection can be such that some of them cannot be discriminated from potential WIMP-induced signals. 

\begin{figure}
\centerline{\includegraphics[width=0.5\textwidth]{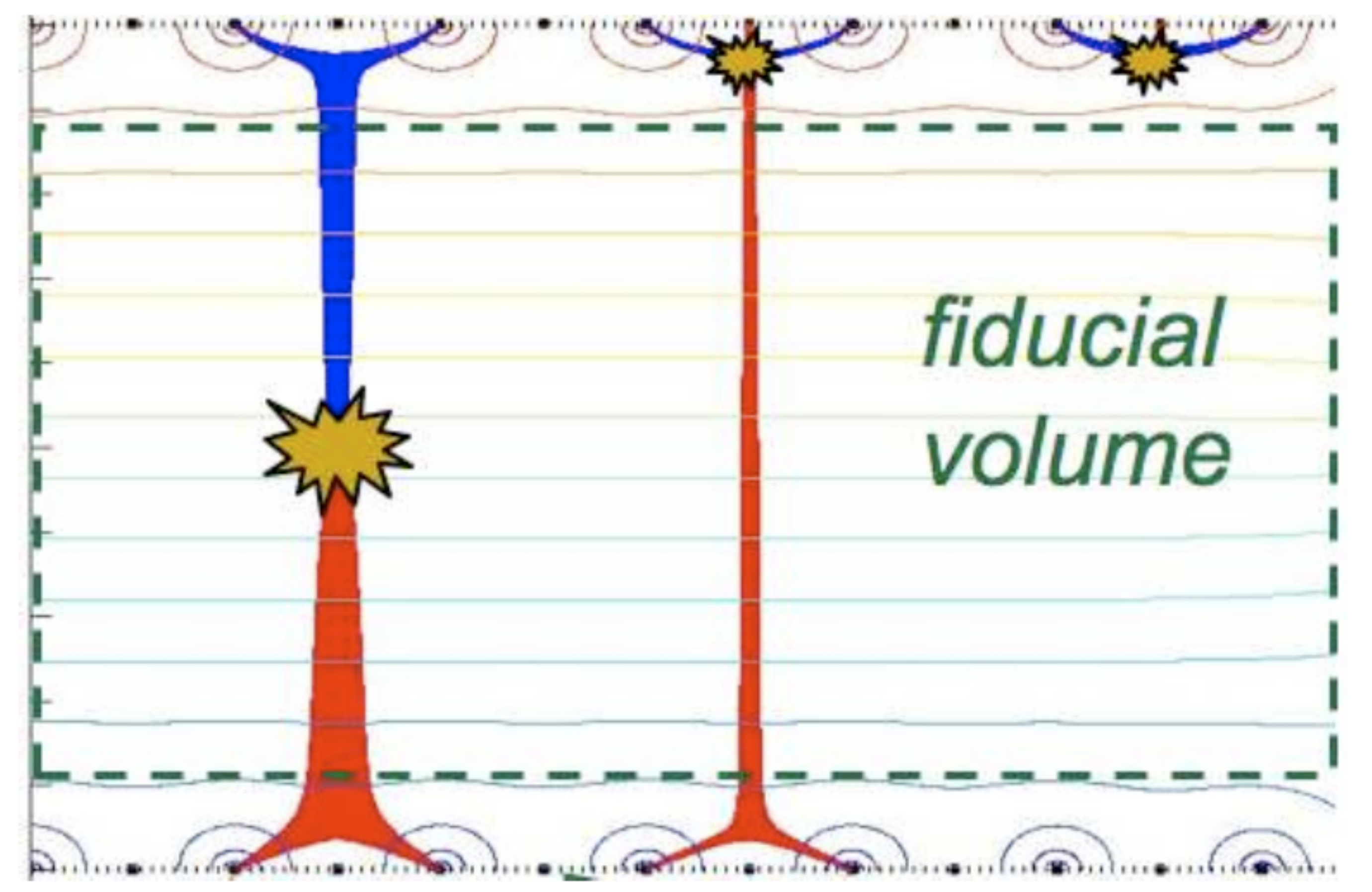}}
\caption{Sectional view illustrating the principle of an ID detector. The interleaved electrodes on the top and bottom of the Germanium crystal create an electric Þeld whose equipotential lines are represented in thin lines. The charge propagation is represented under very simpliÞed assumptions in blue (electrons) and red (holes) for three different interaction positions within the crystal. The distribution of the collected charges in the different electrodes enables to discriminate interactions taking place within the fiducial volume.}\label{Fig:ID}
\end{figure}

In order to reject these events, new-generation detectors called "ID" (InterDigit) were developped. The principle of these detectors is shown on Fig.\ref{Fig:ID} : a set of interleaved electrodes forming concentric rings modiÞes the electric Þeld topology near the crystal surface, and therefore the repartion of charges induced by near-surface events. This allows to tag speciÞcally near-surface interactions and to deÞne a well-controlled Þducial volume for each detector. Note that this strategy to reject surface backgrounds is different from the one used within the CDMS experiment  \cite{CDMS:2009}, which relies on the use of more complex phonon sensors. 

The discrimination against surface events for ID detectors was experimentally proven at a level of 10$^{5}$ using $\beta$-ray sources placed in front of a 200 g ID bolometer \cite{Bronia:2009}. Fiducial interactions are selected down to low energies by requiring both a perfect charge balance between the signals of the top and bottom collecting electrodes,  and the absence of charge deposits on the field shaping electrodes. This dual rejection provides a strong redundance, and enables the detector operation even when charges cannot be read on one of the field shaping electrodes. The rejection quality was found to be independent of the intensity of the applied voltages. 

The fiducial volume of ID detectors may be estimated using a simple electrostatic modelling of the detectors, and has been also measured in real WIMP-search conditions using the 9.0 keV and 10.3 keV $\gamma$-ray lines from the decays of the $^{65}$Zn and $^{68}$Ge isotopes, homogeneously distributed within the crystals due to cosmogenic activation. Comparing the intensity of these lines before and after the fiducial selection provides a measurement of the fiducial mass : for 400 g ID detectors, this mass is 166$\pm$6g. Simulations show that it is primarily determined by the guard regions that are equipped with independent plain electrodes on the edges of the crystals. 

\section{WIMP search with one year data}

A WIMP search was carried out using ten 400 g ID detectors within the EDELWEISS-II setup, mostly between April 2009 and May 2010. Here we present a preliminary analysis of the full data set, using nine out of the ten bolometers, corresponding roughly to a doubling of the exposure with respect to the first 6 months of WIMP search which were already published in \cite{Run12_1:2009} . 
Stable cryogenic conditions at 18 mK were maintained over the whole year without any major interruption. Around 90\% of the electronics channels were working, enabling the use of nine detectors over the ten installed in the cryostat for a first blind WIMP search. External gamma-ray and neutron calibrations were carried out for all detectors. The data was processed using two independent reconstruction pipelines, both of them making use of optimal Þltering in order to adapt to the different noise conditions encountered. The average baseline resolutions were of $\sim$ 1.2 keV FWHM for heat channels and $\sim$ 0.9 keV FWHM for ionizations. Baseline noise measurements are used in order to automatically select good periods of data taking, with an 80\% efficiency. A $\chi^2$  cut is also applied to reject misreconstructed events. WIMPs are then searched among fiducial events, using the ionization yield to discriminate gamma-rays with a 99.99\% rejection efÞciency. In order to reject neutron-induced recoils, events in coincidence between bolometers and also with the muon veto are also rejected. Furthermore, a WIMP search energy threshold was also set a priori at 20 keV in this analysis, so that the search efficiency is independent of the energy. After all cuts, an effective exposure of 322 kg.days is obtained. This analysis procedure is strictly identical to the one published in  \cite{Run12_1:2009}. 

\begin{figure}
\centerline{\includegraphics*[width=1\textwidth]{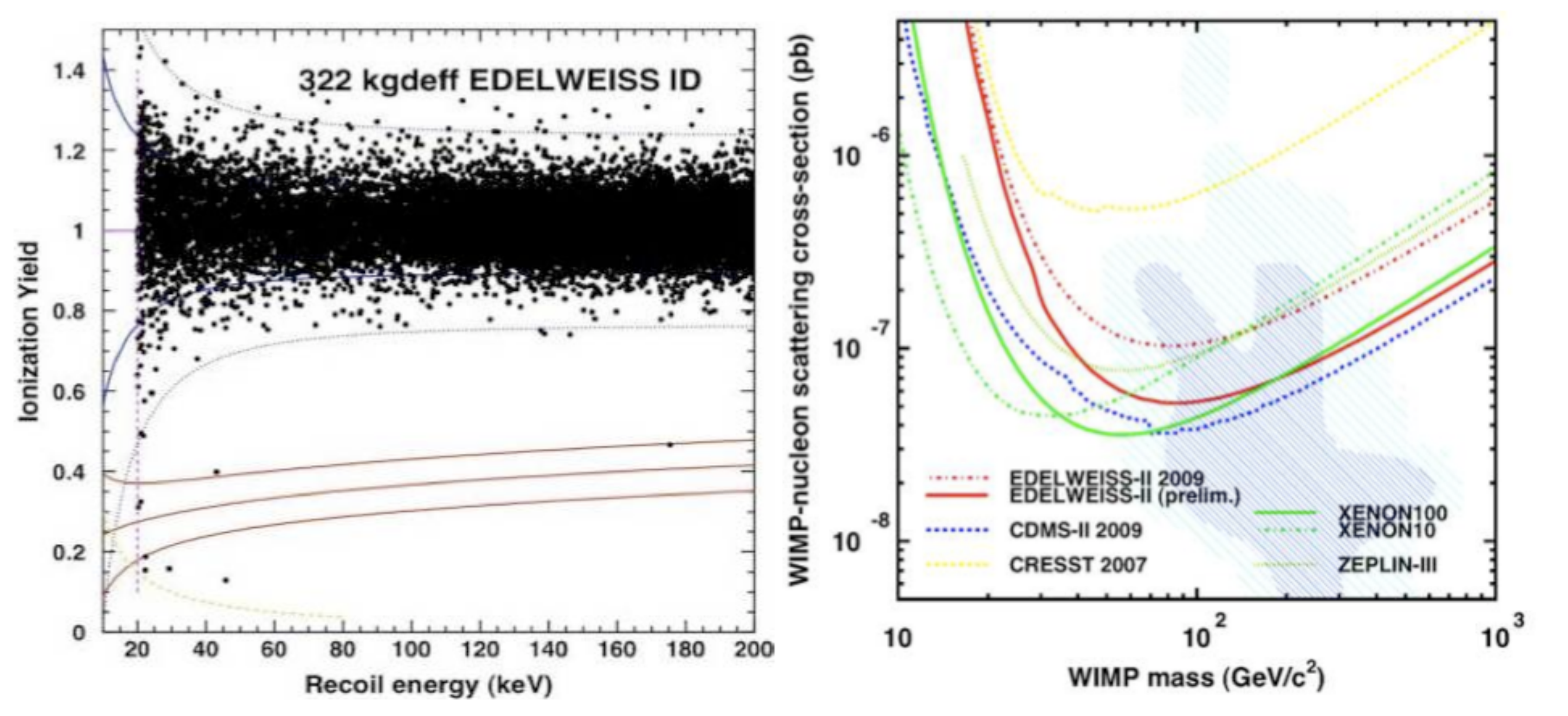}}
\caption{Left : Ionization yield vs energy for all fiducial interactions taking place in a WIMP search corresponding to a 322 kg.days exposure. Right : The corresponding limit on the WIMP-nucleon cross-section (continuous red line) is compared to other experiments.} \label{Fig:lim}
\end{figure}

\begin{figure}
\centerline{\includegraphics*[width=0.6\textwidth]{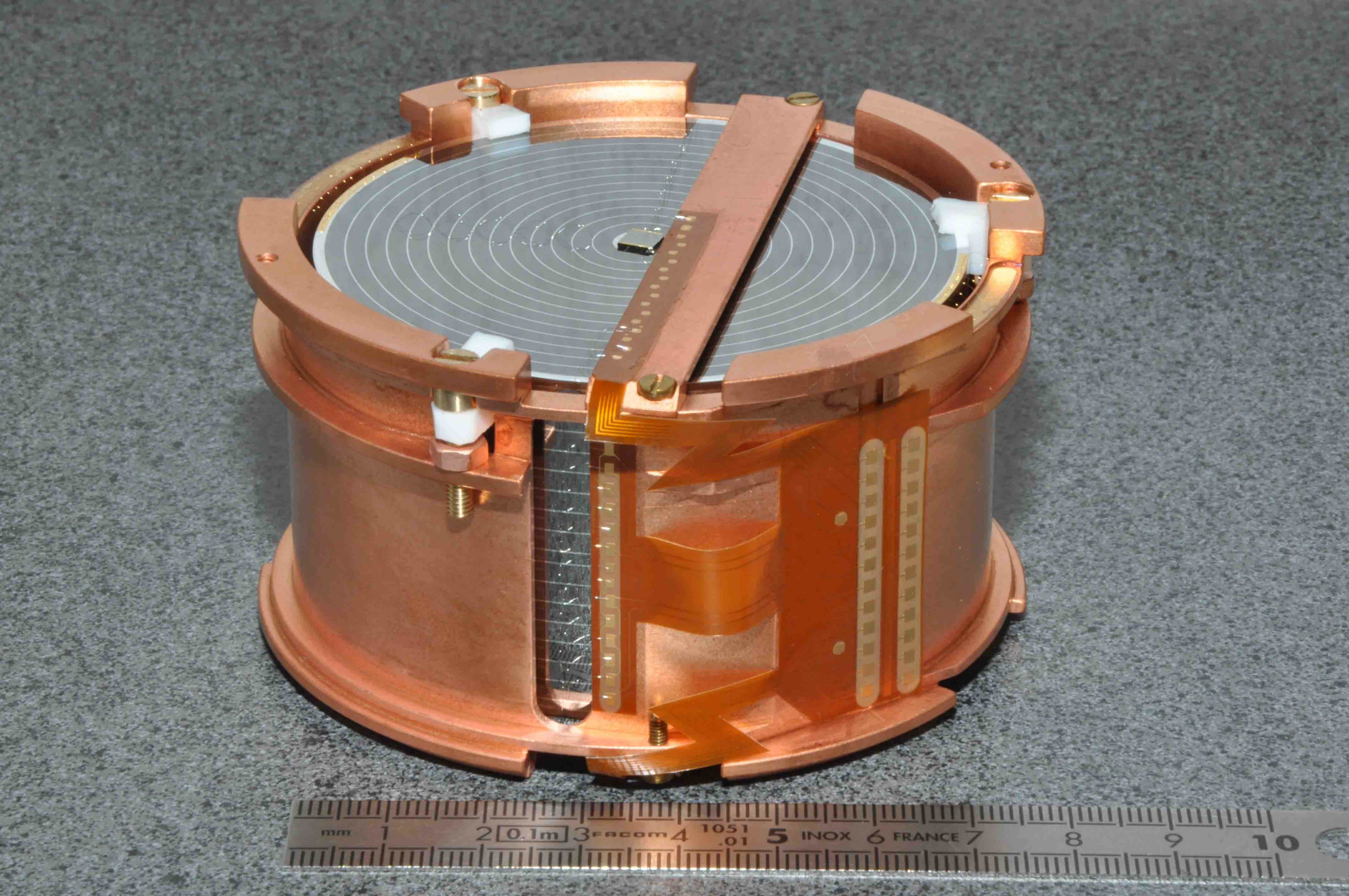}}
\caption{Picture of an 800 g Full InterDigit bolometer. } \label{Fig:fid800}
\end{figure}

The WIMP-gamma discrimination diagram obtained after the Þducial cut is presented on Fig.\ref{Fig:lim}(left). Four events are observed in the band, but with a large gap with no event between 23 and 175 keV. Using the standard Yellin prescription \cite{Yellin:2009}, this enables to improve the EDELWEISS sensitivity by a factor of 2 with respect to the published limit, as shown on Fig.\ref{Fig:lim} (right). A limit on the spin independent WIMP-nucleon cross-section of 5$\times$10$^{-8}$ pb is set for a WIMP mass of 80 GeV. A final analysis of this data set is under way, which aims mostly at optimizing the cumulated exposure using our improved knowledge of these new detectors. 

From the Fig.\ref{Fig:lim}(left), first hints of a residual background start to appear in the data from these detectors : three events are present in the nuclear recoil band near the analysis threshold, between 20 and 23 keV, while two other outliers are present near the nuclear recoil band at higher energies. Note that all these events are well-reconstructed, and comfortably above the noise level of the detectors. Background studies are currently ongoing to fully understand their origin. 

Preliminary upper limits may be derived for the known residual gamma, beta and neutron backgrounds, using calibrations, radioactivity measurements as well as simulations. Overall, less than 1.6 events (90\% CL) of known origin are expected for this WIMP search.

\section{Improved detectors, prospects}
New detectors have been built with interleaved electrodes surrounding the cylindrical part of the crystal, hence their name Full InterDigit (FID). A first series of four 800 g and two 400 g FIDs were built, one of them been shown on Fig.\ref{Fig:fid800}. Using the FID design with an unprecedented mass of 800 g, will increase signiÞcantly the fiducial mass for a given detector. The currently commissionned 800 g detectors at LSM have six charge readout channels and two NTD sensors, enabling an additional redundancy. 
After the validation of these "FID800" detectors, it is planned to upgrade several parts of the EDELWEISS-II setup (shieldings, cryostat, cabling) in order to reduce the gamma and neutron backgrounds and lower the energy thresholds. Then 40 such detectors will be installed, allowing to reach a 3000 kg.days exposure within a few months in 2012, with a potential WIMP-nucleon cross-section sensitivity at the level of 5$\times$10$^{-9}$ pb. 

As a conclusion, the EDELWEISS collaboration has carried out a one-year WIMP search with new-generation ID detectors. The achieved sensitivity of 5$\times$10$^{-8}$  pb for a WIMP mass of 80 GeV is at the same level as the recently published CDMS and XENON100 sensitivities to the standard neutralino-like WIMP models. Optimized detectors are now on the way to be validated in order to improve the current sensitivity by an order of magnitude within the coming years.


\section{Bibliography}


\begin{footnotesize}

\end{footnotesize}



\begin{thebibliography}{99}
\bibitem{Bertone_book:2010}
  K.~Nakamura {\it et al.}  [Particle Data Group], ``Review of particle physics,''
  J.\ Phys.\ G {\bf 37} (2010) 075021, see also ``Particle Dark Matter'', edited by G. Bertone, 2010, 
  Cambridge University Press
\bibitem{Benoit:2001}
 A. Benoit et al., Phys. Lett. B  {\bf513} (2001), 15 [arXiv:astro-ph/0106094]
\bibitem{CDMS:2009}
D.S. Akerib et al., Phys. Rev. Lett. {\bf96} (2006), 011302 [arXiv:astro-ph/0509259]
\bibitem{Bronia:2009}
A. Broniatowski et al., Phys. Lett. B {\bf681} (2009), 305 [arXiv:0905.0753]
\bibitem{Run12_1:2009}
E. Armengaud et al., Phys. Lett. B {\bf687} (2010), 294 [arXiv:0912.0805]
\bibitem{Yellin:2009}
S. Yellin, Phys. Rev. D {\bf66} (2002) 032005
\end{thebibliography}
\end{document}